\documentclass[12pt,preprint]{aastex}
\usepackage{wrapfig}

\newcommand{\css}{CSS~120422}
\newcommand{\cssfull}{CSS~120422:111127+571239}

\slugcomment{Submitted to The Astronomical Journal}

\shorttitle{\cssfull}
\shortauthors{Littlefield et al.}

\begin{document}


\title{A New Sub-Period-Minimum CV with Partial Hydrogen Depletion and Evidence of Spiral Disk Structure}


\author{C.~Littlefield,\altaffilmark{1} P.~Garnavich,\altaffilmark{1} A.~Applegate,\altaffilmark{5} K.~Magno,\altaffilmark{1} R.~Pogge,\altaffilmark{6} J.~Irwin,\altaffilmark{2} G.~H.~Marion,\altaffilmark{2} J.~Vink\'o,\altaffilmark{3,4} R.~Kirshner\altaffilmark{2}}



\altaffiltext{1}{Physics Department, University of Notre Dame, Notre Dame, IN
46556}
\altaffiltext{2}{Harvard-Smithsonian Center for Astrophysics, Cambridge, MA
02138}
\altaffiltext{3}{Department of Optics, University of Szeged, Hungary}
\altaffiltext{4}{Astronomy Department, University of Texas, Austin, TX 78712}
\altaffiltext{5}{Physics Department, Georgian Court University, Lakewood, NJ 08701}
\altaffiltext{6}{Department of Astronomy, Ohio State University, 140 West 18th Avenue, Columbus, OH 43210}


\begin{abstract}
We present time-resolved spectroscopy and photometry of \cssfull\ (= SBS
1108+574), a recently discovered SU~UMa-type dwarf nova whose 55-minute orbital
period is well below the CV period minimum of $\sim$78~minutes. In contrast with most other known CVs, its spectrum features He~I emission of comparable strength to the Balmer lines, implying a hydrogen abundance less than 0.1 of long period CVs---but still at least 10 times higher than than in AM~CVn stars. Together, the short orbital period and remarkable helium-to-hydrogen ratio suggest that mass transfer in \css\ began near the end of the donor star's main-sequence lifetime, meaning that the system is probably an AM~CVn progenitor as described by \citet{phr03}. Moreover, a Doppler tomogram of the H$\alpha$ line reveals two distinct regions of enhanced emission. While one is the result of the stream-disk impact, the other is probably attributable to  spiral disk structure generated when material in the outer disk achieves a 2:1 orbital resonance with respect to the donor.
\end{abstract}


\keywords{stars: cataclysmic variables --- stars: dwarf nova --- stars:
evolution --- stars: individual(\cssfull) --- stars: individual(SBS 1108+574)}



\section{Introduction}

\subsection{Cataclysmic Variables and the Period Minimum} Cataclysmic variables
(CVs) are close binary systems in which a white dwarf accretes matter from a
companion star which overflows its Roche lobe. Very few CVs have orbital periods
below the so-called “period minimum,” which has been observed to be $\sim$78~minutes \citep{hellier01}. As \citet{kolb99} explain, the period minimum is a consequence of emerging electron degeneracy in the donor star and
the strong, inverse relationship between donor-star density and a CV's orbital
period. When a non-degenerate donor sheds some of its mass, its density
increases, prompting the CV's orbital period to shrink. A degenerate donor,
however, has exactly the opposite response to mass loss, so once the secondary
star becomes degenerate, it will have reached its maximum density---and
thus the shortest orbital period possible for the system. Thereafter, the
CV will evolve toward a longer orbital period, its period having ``bounced" off
the minimum.

Nevertheless, a handful of CVs, some with orbital periods as
short as several minutes, defy the 78-minute period minimum. Almost
every CV in this category is an AM~CVn-type system, featuring a rich helium
spectrum lacking even the slightest hint of hydrogen. Helium-enhanced stars are denser than their hydrogen-dominated counterparts, so they adhere to a shorter period
minimum \citep{augusteijn96}.

There are three theoretical avenues of formation for these unusual stars
\citep{nelemans10}. First, the distance between two detached white dwarfs might
decrease as a result of gravitational radiation, causing the lower-mass
companion to eventually overflow its Roche lobe. Second, a low-mass,
helium-fusing star in an interacting binary might lose so much mass that its
core would be unable to sustain the necessary temperature and pressure for
continued helium fusion, producing a semi-degenerate core. Third, if mass
transfer were to commence near the end of the main-sequence lifetime of the
donor star, the secondary star would shed its hydrogen envelope, exposing the
relatively dense, hydrogen-deficient interior of the star. Spectroscopically,
the helium-to-hydrogen ratio would gradually increase as the system continued to
evolve. Unlike the first two models, the evolved-main-sequence-donor model
predicts the presence of some hydrogen in the donor star \citep{phr03,
breedt12}.

Oddly, a few CVs are below the period minimum, but with strong hydrogen lines in
their spectra, they cannot be AM~CVn stars. Instead, these stars belong to the
SU~UMa sequence of CVs; consequently, while in outburst, they display
superhumps---periodic photometric oscillations attributed to a precessing,
eccentric accretion disk---whose period is within several percent of the orbital
period. However, the secondary stars in the sub-period-minimum SU~UMa systems
are not the garden-variety donor stars found in most CVs above the minimum. In
two of these CVs, EI Psc \citep{thorstensen02} and V485 Cen
\citep{augusteijn96}, the He~I 667.8-nm-to-H$\alpha$ ratio is elevated with
respect to CVs above the period minimum, and the secondary star is plainly
visible in optical and near-infrared spectra. A third system below the minimum,
SDSS J150722.30+523039.8, has very weak helium lines and shows no trace of the
donor, leading \citet{littlefair07} to conclude that the secondary is a brown
dwarf.

Another sub-period-minimum star, CSS 100603:112253-111037 (hereinafter,
CSS~100603), is a hybrid of these three systems. Like EI~Psc and V485~Cen, its
spectrum has strong emission from both hydrogen and neutral helium.
Like the brown-dwarf system, though, the donor star is invisible in spectra, and
the estimated mass ratio is quite low. \citet{breedt12} argue that CSS~100603 is
probably an AM~CVn progenitor in which the donor star is an evolved
main-sequence star.

Here, we present observations of \css, a very short period system similar to CSS 100603, but with several interesting differences.

\subsection{\cssfull}

On 2012 April 22, the Catalina Sky Survey \citep{drake09} detected an outburst
of \cssfull\ (hereinafter, \css), a previously undiscovered CV. With a peak
observed brightness of V $\approx$ 15.4, the system brightened by at least 4
magnitudes from its typical brightness in the Catalina data.\footnote{Because
\css\ was discovered at an indeterminate point after the start of its outburst, the
actual peak magnitude is unknown.} Even prior to the discovery of its very
short orbital period, the system garnered attention because it is remarkably
blue during quiescence, even for a CV \citep{kato_outburst}. The VSNET
collaboration \citep{kato_vsnet} immediately launched an extensive observing
campaign, finding an initial superhump period of 56 minutes, well below the
78-minute period minimum for hydrogen-rich CVs \citep{kato_astroph}. Although
most known CVs below the minimum are AM CVn systems, spectroscopy revealed a
hydrogen-dominated spectrum, eliminating the AM CVn hypothesis
\citep{garnavich12}. After the system returned to quiescence, Pavlenko et al.
\citep[in preparation; preliminary findings discussed in][]{kato_astroph}
reported unusually strong helium lines in \css's quiescent spectrum.

\section{Observations}

\subsection{VATT Photometry}

We observed \css\ with the Vatican Advanced Technology Telescope (VATT) and
VATT4K CCD imager on 2012 June 15, 16 and 18 (UT), performing time-resolved
photometry through a Bessell $B$ filter on the first two nights a Bessell $R$
filter on the final night. Clouds prevented observations on June 17. The CCD
pixels were binned 2$\times$2, and only the central 500 rows were read out in
order to reduce the overhead on each exposure. Generally exposures were 60~sec
long with a readout time of 11~sec, but seeing degraded during the second half
of the final night, so the exposure time was increased to 100~sec. The CCD images
were bias-subtracted and flat-field-corrected using dithered exposures taken
during twilight.

\begin{figure} [ht!]
\epsscale{.85}
\plotone{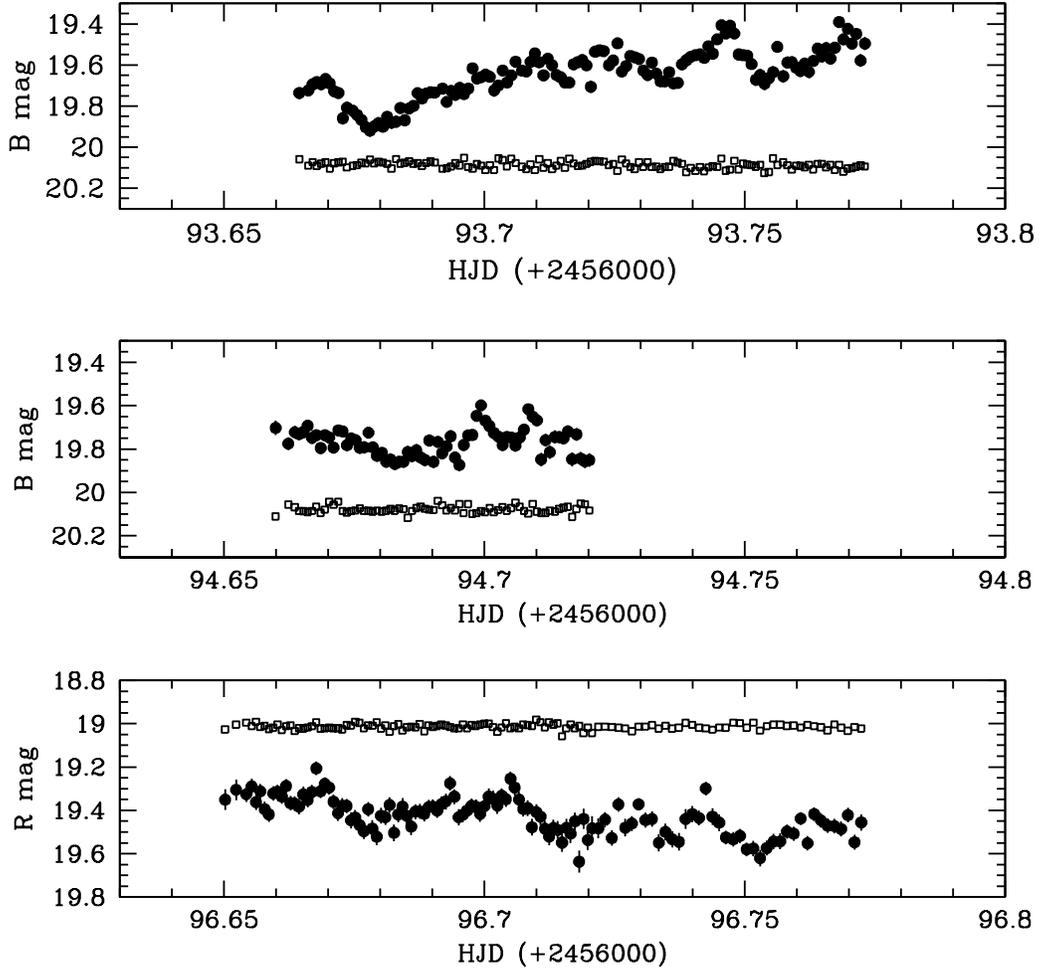}
\caption{ \it{The VATT light curves of \css\ obtained in 2012 June. The open
circles show the light curve of the check star, offset by adding 1.0~mag. The waveform is highly variable, but the double-peaked structure visible in the phase plot in Fig.~\ref{vatt} is intermittently visible. Flickering is evident in all three light curves.} \label{vatt_phot}}
\end{figure}

We used a star 2~arcmin south of \css\ as a comparison star (USNO-B1.0 position
RA=11:11:26.83 DEC=+57:12:38.9 J2000). The SDSS photometric catalog magnitudes
for the comparison star convert to $B=16.23\pm 0.05$ and $R=14.62\pm 0.04$ mag.
Located 1~arcmin west of the comparison star, the check star has SDSS converted
magnitudes of $B=19.1$ and $R=18.0$. Aperture photometry using DAOPHOT in IRAF produced the light curves shown in Fig.~\ref{vatt_phot}.

\subsection{FLWO Spectroscopy}

We obtained spectra of \css\ with the 1.5m Tillinghast telescope at the Fred
L. Whipple Observatory on 2012 May~12 and 13 (UT), roughly three weeks after the
initial detection of the outburst. The FAST spectrograph utilized a
300~line~mm$^{-1}$ grating and covered the spectral range between 360~nm and
750~nm. On May~12, a single 30~minute exposure with a 3-arcsec-wide slit was
obtained. During the following night, a series of nine images, each with an
exposure time of 420~sec, were taken through a 2-arcsec-wide slit. Seeing was
approximately 1.5~arcsec, and the slit was set to a position angle of 90$^\circ$
on both nights. Table~\ref{speclog} gives a log of the spectral observations.

The CCD images were bias-subtracted and flat-field-corrected using normalized
internal lamps. Extracted using the ``twod" package in
IRAF,\footnote{IRAF is distributed by the National Optical Astronomy
Observatory, which is operated by the Association of Universities for Research
in Astronomy (AURA) under cooperative agreement with the National Science
Foundation.} the spectra were wavelength-calibrated using images of an internal
HeNeAr lamp interspersed with the stellar spectra. The \css\ spectra, having all been obtained at an airmass less than 1.15, were flux-calibrated using a spectrum of Feige~34 taken at an airmass of 1.03.

\begin{figure}[ht!]
\epsscale{.65}
\plotone{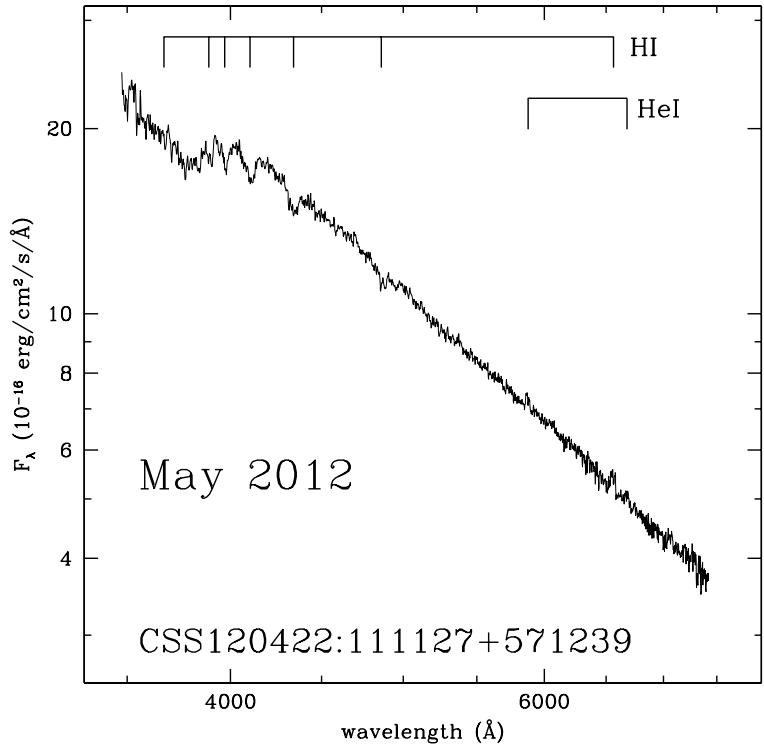}
\caption{\it{The FLWO spectrum taken in 2012 May when the star
was declining from outburst. Weak H$\alpha$, He~I 587.5~nm, and He~I 667.8~nm emission lines are visible, and H$\beta$ shows a combination of emission and absorption. Absorption dominates the higher-level Balmer lines.}
 \label{spec_may}}
\vspace{-0.5cm}
\end{figure}

The combined spectrum, displayed in Figure~\ref{spec_may}, reveals a blue
continuum with a conspicuous Balmer jump. Weak emission lines of H$\alpha$ and
He~I at both 587.5~nm and 667.8~nm are visible in addition to broad Balmer
absorption features with emission cores, a spectrum typical of a
hydrogen-accreting dwarf nova declining from outburst \citep{garnavich12}. Because the system was returning to quiescence, its absorption-dominated outburst spectrum was in the process of transitioning back into an emission-line spectrum during our observations. Consequently, the simultaneous presence of emission and absorption undoubtedly diluted the strengths of many lines.

\subsection{LBT Spectroscopy}

We obtained spectra of \css\ with the Large Binocular Telescope (LBT) and
Multi-Object Dual Spectrograph \citep[MODS;][]{pogge10} on 2012 June~16 (UT).
The sixteen 400-second exposures employed a 1.0~arcsec slit oriented to the
parallactic angle in order to minimize slit losses. The spectra cover a time
period of about 1 hour and 15 minutes, or about 1.3 orbits of the binary.
Table~\ref{speclog} presents a log of the LBT spectral observations.

The red and blue MODS CCD spectra were bias-subtracted and flat-field-corrected
using normalized internal lamp illumination images. As with the FLWO data, the
spectra were extracted using the ``twod" package in IRAF. We used an internal
Neon lamp on the red side and an Argon lamp for the blue channel for wavelength
calibration, and the star HZ~44 served as the spectrophotometric standard for
the flux calibration.

The brightness of \css\ in the LBT data is approximately 2.8~mag fainter than
the FLWO spectra obtained in May.

\begin{figure} [ht!]
\epsscale{.80}
\plotone{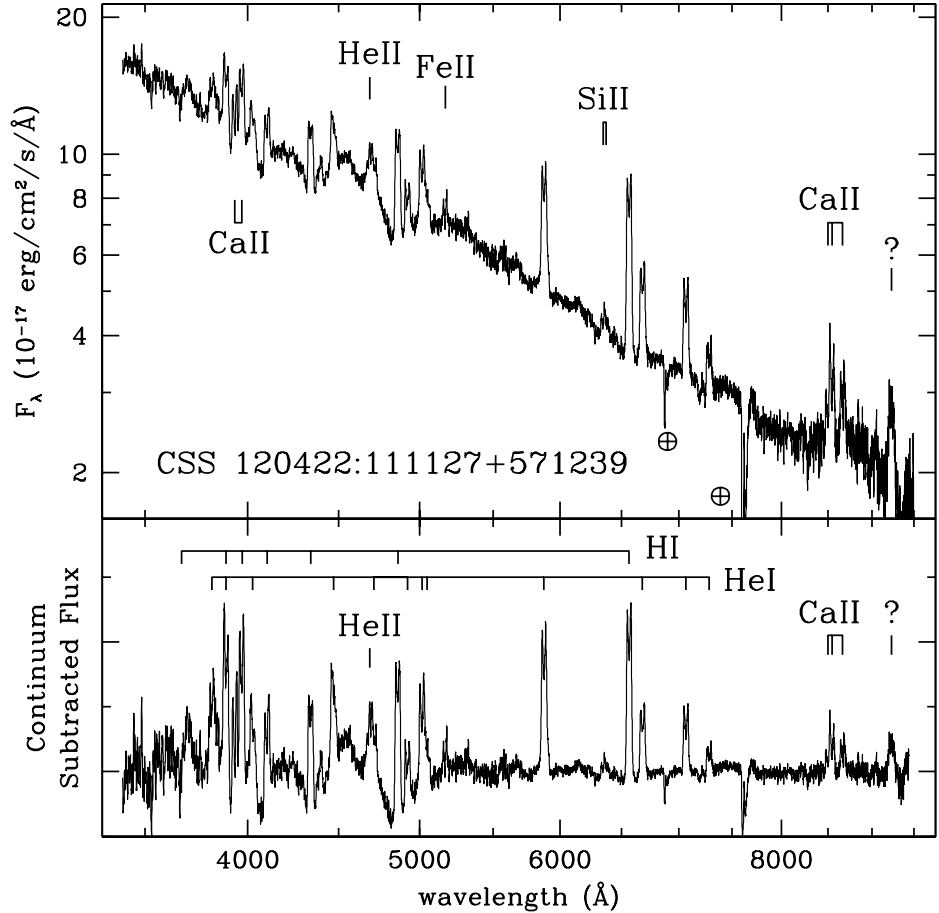}
\caption{ \it{The LBT spectrum of \css\ obtained in 2012 June after the system had faded from its outburst. The spectrum is the sum of sixteen 420-second exposures.  In the bottom panel, the strong continuum which dominates the top panel has been subtracted. The simultaneous presence of strong Balmer and He~I emission is an immediate indicator that the system is unusual. Ca~II and Si~II emission features are also conspicuous. The weak, highly phase-dependent He~II emission at 468.6~nm is partially blended with the He~I 471.3~nm line.}
 \label{spec}}
\end{figure}

The combined LBT spectrum appears in Figure~\ref{spec}. Similar to the May spectra,
a strong blue continuum is obvious, but by June, a wide variety of emission lines
had appeared. Double-peaked Balmer and He~I emission lines which
originate in the accretion disk are evident, as are weaker lines of
intermediate-mass elements, especially Si~II and Ca~II. The H$\alpha$ profile is
well fit by two Gaussians separated by 1200~km~s$^{-1}$, each with a full width
at half maximum (FWHM) of 940~km~s$^{-1}$. The center of mass of the line is
shifted to the red by 50~km~s$^{-1}$.

The near-equal strength of the He~I and Balmer lines is particularly striking. In normal, hydrogen-rich CVs, the He~I emission is much weaker than the hydrogen lines, and in AM CVn systems, He~I dwarfs the negligible hydrogen emission. However, in \css, the average total flux of the He~I 587.5~nm line is 20\%\ higher than that of H$\beta$. To put this in perspective, the H$\beta$ line is typically three times as strong as the He~I 587.5-nm feature in CVs.

\begin{figure}[ht!]
\epsscale{.75}
\plotone{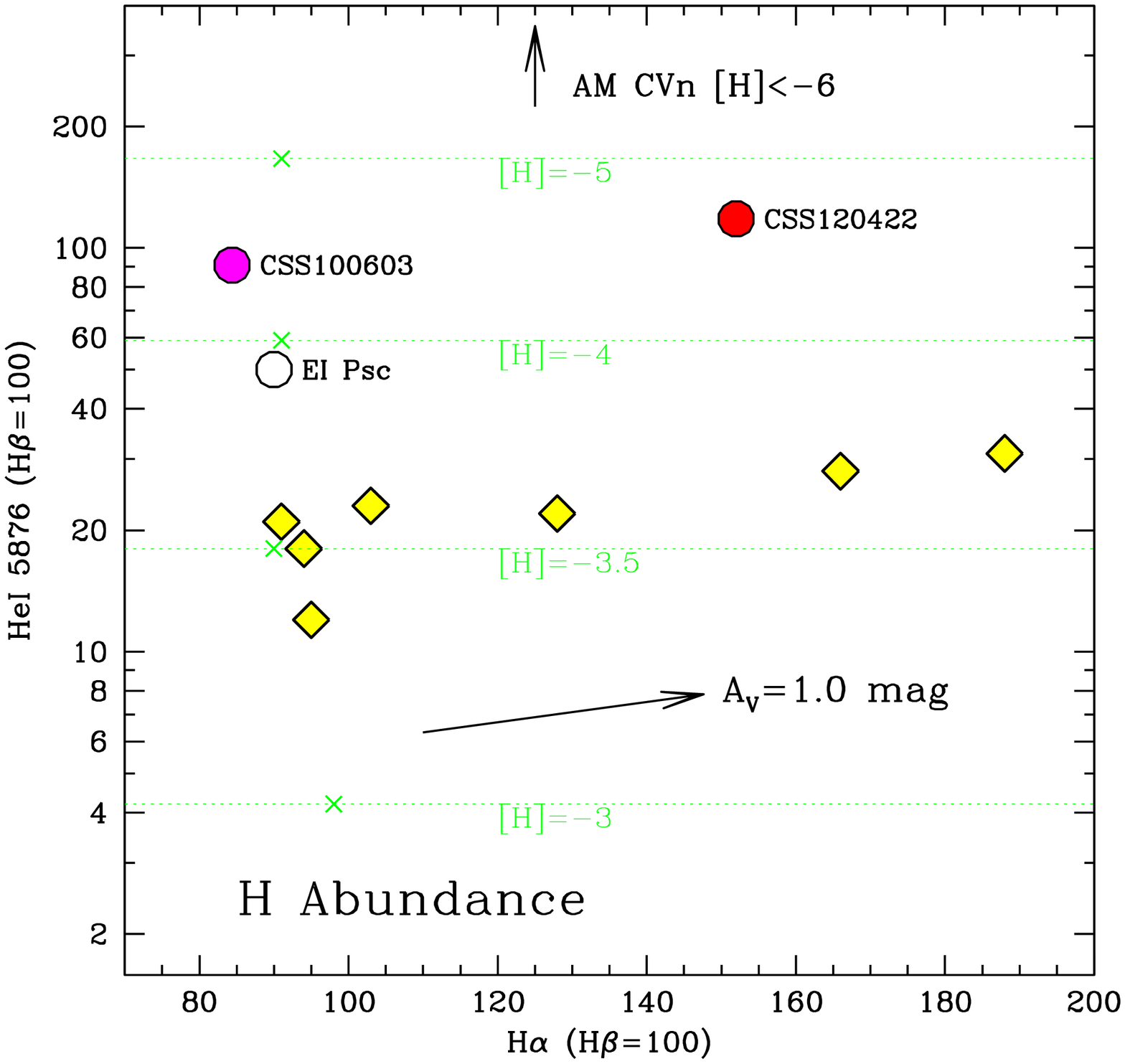}
\caption{ \it{The He~I 587.6-nm line flux versus the H$\alpha$ flux relative to
the strength of H$\beta$. The diamonds show the long-period CVs studied by
\citet{williams82}, while the circles show new, short-period CVs. No correction
has been made for reddening, and the affect of one magnitude of visual
extinction is indicated by an arrow. The He~I 587.6-nm line strength is
estimated from the flux of the 667.8-nm line. The dotted lines indicate hydrogen
abundance relative to Solar based on \citet{williams82} and \citet{nagel09}
models. The ``x" on each dotted line shows where the H$\alpha$ flux is predicted
by \citet{williams82}.}
 \label{ratio}}
\end{figure}

As alluded to earlier, other systems, such as CSS~100603 and EI~Psc, exhibit
He~I-to-H ratios comparable to that of \css. \citet{williams82} modeled the line
strengths in several long-period CVs and found that they were somewhat depleted
in hydrogen compared to the Sun. Specifically, the He~I 5876-to-H$\beta$ flux
ratio observed in long-period CVs was about 0.2$\pm0.1$, implying a hydrogen
abundance [H]=log$_{10}$(H/He)/Solar(H/He)=$-3.5$ according to
\citet{williams82}. \citet{nagel09} concluded that significant H$\alpha$
emission lines are present at [H]=$-5$ and that AM~CVn stars must have [H]$<-6$
for hydrogen emission to be undetectable (see Fig. \ref{eqw}). With some
interpolation, these studies indicate that in \css, which features a He~I
587.6-nm line of similar strength to the H$\alpha$ line, the hydrogen abundance
is [H]$\approx -4.5$. We also measured this line ratio for CSS~100603 from the
public SDSS spectrum and find a hydrogen abundance that is similar to \css. The
He~I 587.6-nm line was not measured by Thorstensen et al., but the He~I 667.8-nm
line strength suggests that EI~Psc has a hydrogen abundance which is a factor of
three higher in comparison with these recently discovered objects.

In \css, Ca~II emission from the infrared triplet and HK lines is also quite prominent.
Additionally, we identify Si~II 634.7/637.1~nm lines, both of which have been
observed in the AM~CVn star CP~Eri \citep{groot01}. The feature at 923~nm may be
due to Mg~II, but the line expected at 788.5 is not detected \citep{marsh91}.
Alternatively, it might be a Paschen line. There is also a weak He~II 468.6~nm feature blended with the He~I 471.3~nm line, and on the red side of the He~I 587.5~nm line,  there is weak emission which may be due to low-velocity Na~I.

\css's spectrum contains more than just this forest of emission lines. In
particular, the H$\beta$ and H$\gamma$ lines both sit within broad absorption
troughs in the continuum. (To a lesser extent, this is also true of the
H$\alpha$ line, too.) In other short-period CVs, the WD is responsible for
producing comparable features at these wavelengths \citep{rodriguez-gil05}, so
we attribute the absorption in \css\ to pressure-broadened absorption from the
WD's photosphere. These absorption lines suggest that the WD must contribute
much of the system's overall optical flux \citep[e.g.][]{patterson08},
likely indicating a low mass-transfer rate during quiescence. If this
inference is correct, it would account for the unusually blue quiescent color
noted by \citet{kato_outburst}. A quiescent accretion disk will usually be
cooler than the WD, causing the overall flux of the system to be redder, but in
a system with a low rate of mass transfer, the disk will be relatively
tenuous and dim, enabling the WD's blue continuum to overwhelm the disk's meager
contribution.

\section{Analysis}

\subsection{VATT Photometry}

\subsubsection{Orbital Period, Ephemeris, \& Light Curve Morphology } 
\begin{figure}[ht!]
\epsscale{1.1}
\plottwo{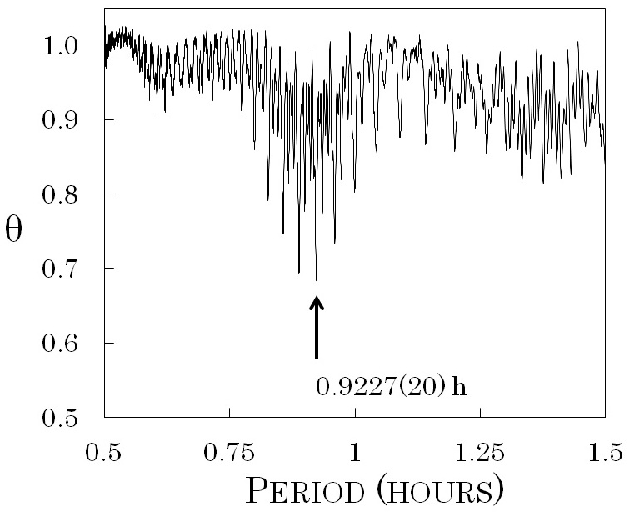} {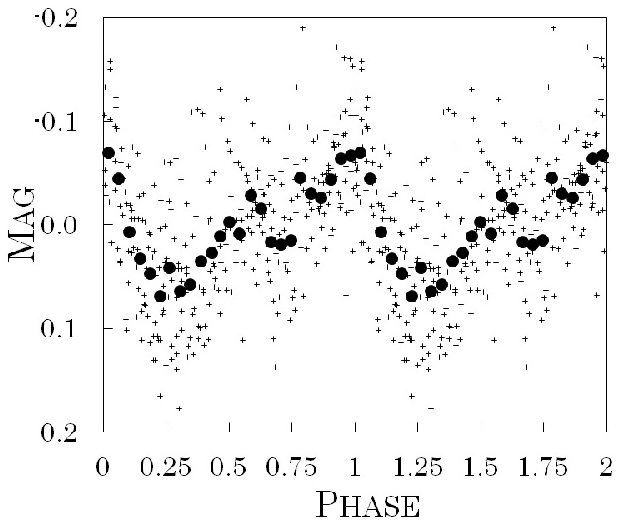}
\caption{\it{{\bf Left:} Power spectrum of three nights of VATT photometry using the phase-dispersion-minimization algorithm. Candidate periods appear as local minima. {\bf Right:} Phase plot of the VATT photometry using the 0.9227-hour period. The dark circles are 0.04-phase bins. The waveform is weakly double-peaked, consistent with theoretical models of CVs with spiral shocks at the 2:1 resonance in the disk (see text).  As Figure \ref{vatt_phot} illustrates, there is considerable variation in the light curve from orbit-to-orbit.}
 \label{vatt}}
\end{figure}

We removed a linear trend from each night of VATT photometry and created a power
spectrum from all three nights using the phase-dispersion-minimization (PDM)
algorithm \citep{stellingwerf78}. PDM scans a user-selected period range,
calculates a phase plot for each trial period, divides the phase plot into a
user-specified number of bins, and quantifies the total amount of scatter in
each bin. The best candidate periods will have the least amount of scatter in
their phase plots, a technique which is well-suited to analyzing non-sinusoidal
signals, such as the waveform of \css's orbital period. In the PDM power
spectrum (see Fig. \ref{vatt}), the strongest signal is at 55.36$\pm0.13$ minutes, but there is also a competing alias at 53.28$\pm0.12$ minutes. Using a completely independent dataset, \citet{kato_astroph} report the longer of these two periods as the orbital period and find no evidence of the shorter period (T. Kato, private
communication).

We show the phase plot of all three nights of data (both $B$ and $R$ bands) in Figure~\ref{vatt} using the ephemeris of $$T_{max}(BJD)\; =\; 2456093.746+0.03845(2)E,$$ where phase zero corresponds with the peak in the light curve.

The phased light curve is very saw-toothed, with a gradual rise, a sharp peak,
and then a rapid drop in brightness. When the data are binned to improve the
signal-to-noise ratio, the waveform displays a weakly double-humped structure,
with two distinct maxima per orbital cycle. In this regard, the light curve
resembles that of WZ~Sge-type CVs. The maximum at phase zero and the broad minimum centered on phase $\sim$0.25 are both readily apparent, but the maximum seen near phase $\sim$0.55 and the ensuing minimum around phase $\sim$0.7 are much more subtle. The phase plot generated by Kato et al. (in preparation) is generally similar, but the double-peaked structure is somewhat difficult to discern.

The morphologies of the individual light curves are remarkably variable from
night-to-night and even from cycle-to-cycle. For example, the light curve from
2012 June 15 should have contained three minima similar to the one centered on
photometric phase $\sim$0.2 in the phase plot in Figure~\ref{vatt}, but
the second one is nearly non-existent. Likewise, on the following night, an
expected peak in the light curve did not occur and was followed instead by a
comparatively gentle decrease in flux. In the very next cycle, however, the peak
returned, and the post-peak dip had a different shape. Clearly, the source of
the orbital modulation, which we believe to be tidally-induced spiral structure
in the outer disk (see Section~\ref{spirals}), is capable of significant
variability on timescales of less than one hour.

\subsubsection{Inferred Mass Ratio} \citet{patterson98} reported an empirical
relation between a CV's mass ratio ($q$ = $m_{2}$/$m_{1}$) and its fractional
superhump period excess ($\epsilon = [P_{sh}/P_{orb}]-1$). Although the exact
physical processes underlying this $\epsilon$-$q$ correlation are unclear, it
provides approximate estimates of $q$ using just the orbital and superhump
periods, making it a particularly useful technique for analyzing a system whose
mass ratio cannot be ascertained by other means. Relying upon a detailed survey
of superhumps, \citet{kato09} updated Patterson's original formula slightly,
obtaining $\epsilon = 0.16q+0.25q^{2},$ where $\epsilon$ is determined
using the shortest P$_{sh}$ rather than the mean P$_{sh}$. Applying the
quadratic formula and rejecting the negative root (to ensure a positive value of
$q$) yields a more convenient, explicit function: $q\;=\;-0.32+2\sqrt{0.0256+\epsilon}.$ Using this formula, a P$_{sh}$ of 55.971~minutes \citep{kato_astroph}, and our P$_{orb}$ of 55.36~minutes, we obtain $q~=~0.06\pm0.01$. However, since the $\epsilon$-$q$ relation in other systems is only approximate, the uncertainty for $q$ is probably greatly underestimated.

The correlation between $\epsilon$ and $q$ underscores that the 53-minute signal
in our photometry is an alias of the true period at 55~minutes. If the signal at 53~minutes were the actual orbital period, then $q~=~0.24$, a value which would be difficult to reconcile with the complete non-detection of the donor and the
low mass ratios of other short-period CVs. Additionally, it would be
inconsistent with the evidence of spiral shocks produced by disk material at the
2:1 resonance which, as Section~\ref{spirals} explains, could form only if $q \lesssim 0.1$.

\citet{kato_astroph} point out that a mass ratio of 0.06 is not nearly
as extreme as the mass ratios of other short-period dwarf novae,
suggesting an unusually dense and massive secondary star which has shed much of
its hydrogen envelope via mass transfer. In Section~\ref{evolution},
we reach a similar conclusion based on our interpretation of the system's
spectrum.

\subsection{Spectroscopy}

The double-peaked Balmer and He~I emission lines in the LBT spectra exhibit a
radial-velocity variation over the photometric period (Figure~\ref{halpha_vel}).
In contrast to many AM~CVn stars and CSS~100603, these lines do not show
stationary, central peaks in our trailed spectra, but there is an `S-wave'
present between the peaks of each of these lines. The
single-peaked He~II emission shows a particularly prominent S-wave for approximately half of the orbit, a feature which we attribute to the bright spot created by the stream-disk collision. The non-appearance of the He~II line for over half of the orbit implies that the inclination of the system is sufficiently high that the bright spot rotates behind the disk as seen from Earth. 

\subsubsection{Equivalent Width Variation}

\begin{wrapfigure}{r}{10cm}
\vspace{-0.5cm}
\epsscale{.6}
\plotone{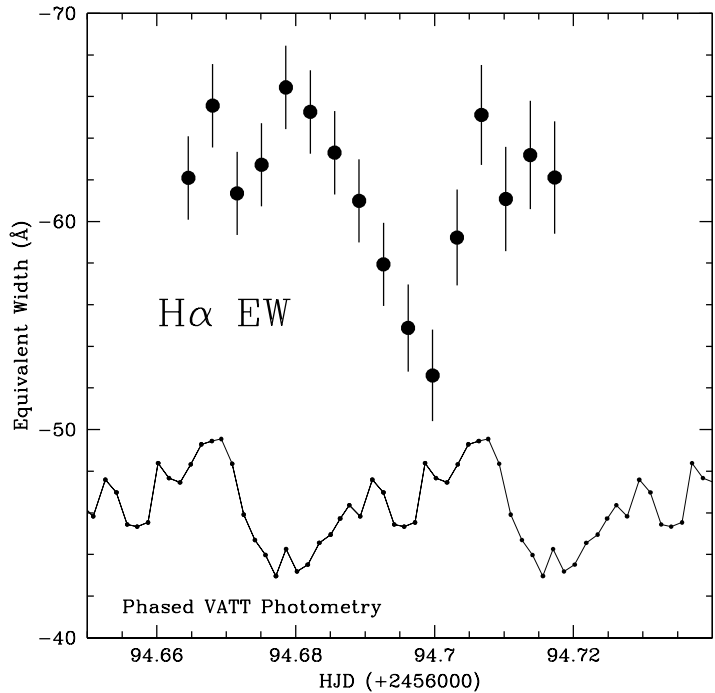}
\caption{\it{The H$\alpha$ equivalent width as a function of time from
the LBT spectra. Over-plotted at the bottom is the photometric phase plot
from the VATT observations.}
 \label{eqw}}
\end{wrapfigure}

We estimated the equivalent width (EW) of several of the brightest lines using
the ``splot" routine in IRAF. Each line was measured several times with slightly
different continuum values and the resulting differences used to estimate an
error. Only H$\alpha$ showed significant variation above the noise, and the
result is shown in Figure~\ref{eqw} along with the phase plot. The H$\alpha$ EW
varies by 20\%, with the lowest EW occurring at photometric phase $\sim$0.8,
when the orbital modulation nears its peak flux.

\subsubsection{Radial Velocities}

To avoid influencing the measurement of radial velocity variations with changes
in the line shapes, we estimated the line velocities using the method described
by \citet{shafter83}. The Shafter method multiples the red and blue wings of an
emission line by a pair of Gaussian functions separated by a certain number of \AA ngstroms. The wavelength of the Gaussians are shifted until the difference in flux measured by the two functions is zero. This wavelength is assumed to be the central
wavelength of the emission line. This is repeated for all the spectra, at which
point it is possible to construct a radial velocity curve. The best separation
between the Gaussian functions is determined by fitting each set of radial
velocity curves with a sinusoidal function and choosing the one that gives the
lowest $\chi^2$ parameter.

\begin{wrapfigure}{r}{10cm}
\epsscale{.6}
\vspace{-0.2cm}
\plotone{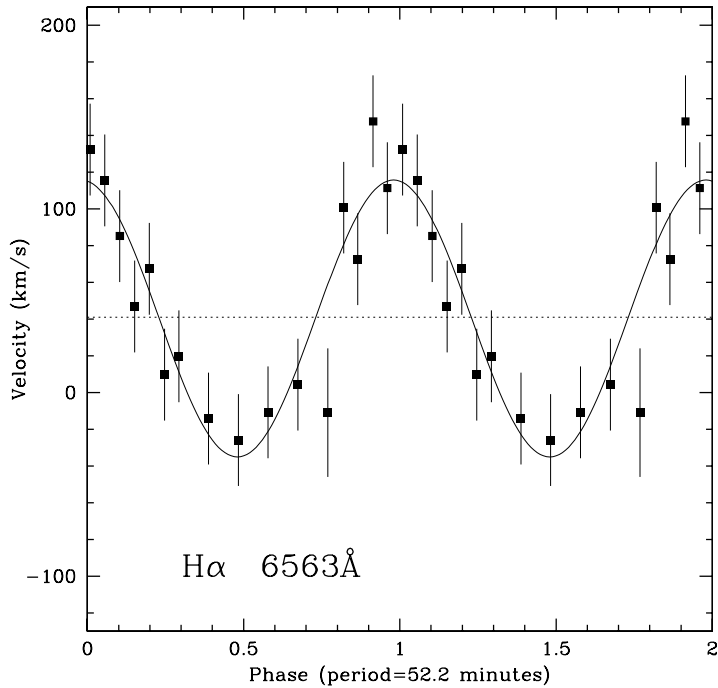}
\caption{\it{The radial velocity derived from the H$\alpha$ line
plotted against phase for a period of 52.2~minutes. The data are
shown from the full Gaussian separation of 61~\AA.}
 \label{halpha_vel}}
\vspace{-0.5cm}
\end{wrapfigure}

\begin{figure} [ht!]
\epsscale{.7} 
\plotone{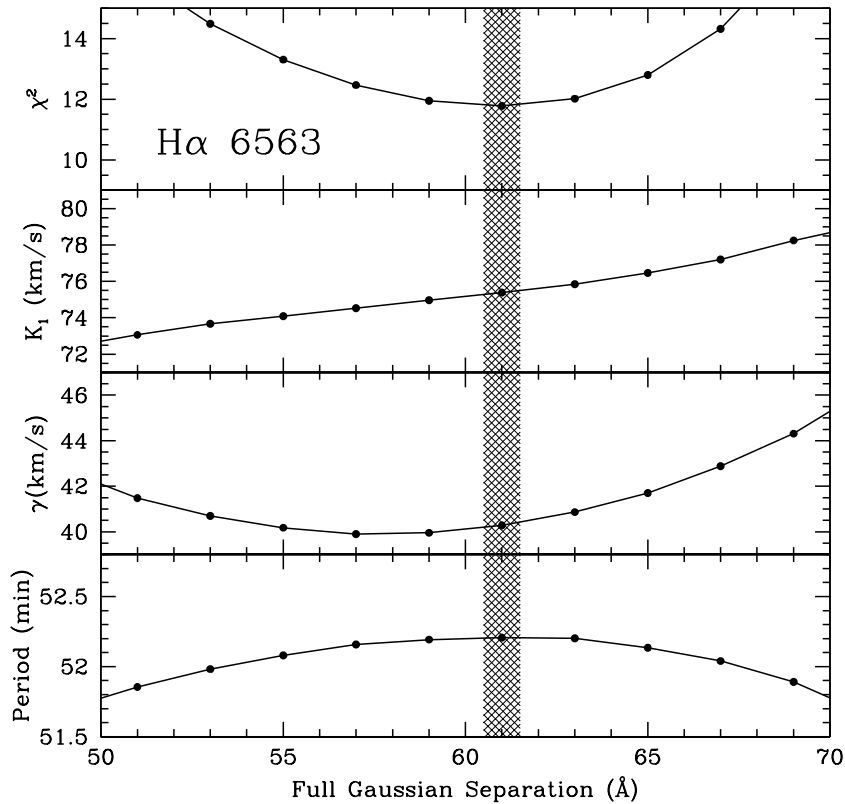} 
\caption{\it{The Shafter fit parameters as a function of the full separation between the Gaussian functions for the H$\alpha$ emission line. The top panel shows the $\chi^2$ parameter and a shaded region indicates the minimum at 61~\AA. The second panel from the top shows the velocity amplitude with the best fit at 75~km~s$^{-1}$. The third panel from top displays the systemic velocity, which has a value of 41~ km~s$^{-1}$ at the minimum $\chi^2$ parameter. The bottom panel shows the orbital period, which never exceeds 52.2~min.}
 \label{ha_shafter}} 
\end{figure}

This method was applied to the H$\alpha$ line and the fit parameters as a
function of Gaussian separation are shown in Figure~\ref{ha_shafter}. The
minimum $\chi^2$ occurred at a full Gaussian separation of 61~\AA, and the
$\chi^2$ parameter rises sharply when the separation is increased. The minimum
$\chi^2$ was 12 for sixteen spectra and four fit parameters (amplitude, $K_1$; systemic velocity, $\gamma$; orbital period, $P$; phase, $\phi$). We find that $K_1=75\pm4$ km s$^{-1}$, $\gamma = 41\pm3$ km s$^{-1}$, and $P = 52.2\pm2.3$ min, a value within 1.4$\sigma$ of the 55.36-minute photometric period.

At first, our estimate of $K_1=75\pm4$ km s$^{-1}$ might seem to contradict the small mass ratio that we have inferred, as it is impossible to obtain so large a value for the WD's radial-velocity amplitude using $q = 0.06$. Nevertheless, $K_1$ in a CV can often be significantly different than the true orbital motion of the WD \citep [e.g. U Geminorum:] [] {lp99}. Since an accretion disk's emission is often asymmetric, techniques which rely upon a measurement of the wings of a spectral line will frequently produce inaccurate radial velocities for the WD \citep[e.g. Appendix A.2 in][]{hellier01}. For example, although CSS~100603 has a mass ratio of  just $q = 0.017$, $K_1$ in that system is $69.4\pm2.9$ km s$^{-1}$ for the H$\alpha$ line, a number which would be unreasonably high if it actually represented the WD's motion.

The resulting radial velocity curve is shown in Figure~\ref{halpha_vel} where
the phase has been set to match the photometric phase. The maximum brightness in
the orbital modulation occurs at the maximum disk redshift.

\section{Discussion}

\subsection{Evolutionary Track of \css}\label{evolution}

Several attributes of \css\ distinguish it from most other sub-period-minimum
systems with detectable hydrogen. For example, EI~Psc and V485~Cen both contain
prominent secondary stars which are relatively easy to detect spectroscopically.
The prevailing explanation for the secondary stars in these two systems is that
they are evolved stars which have shed their outer envelopes. Both are
abnormally luminous for such short orbital periods \citep{uemura02}. While
EI~Psc and V485~Cen both exhibit helium enrichment like \css, the non-detection
of the donor in \css\ out to 950~nm suggests that it is cool and dim, in
agreement with the system's very low mass ratio.

While the combination of an invisible donor and a sub-minimum period could be the
signature of a brown-dwarf secondary, the greatly elevated levels of helium in
\css\ do not support this scenario. Brown dwarfs, by definition, lack the mass
necessary to enter the main sequence, so they should not be particularly rich in
helium. Thus, the helium enhancement in EI~Psc and V485~Cen is inconsistent with the presence of brown-dwarf donors in those systems \citep{politano}; indeed, in the spectra of brown-dwarf CVs, the He~I emission is quite subdued \citep[e.g.,][]{littlefair07, unda-sanzana08}. Although the He~I/H$\alpha$ ratio is sensitive to both pressure and temperature, the intense He~I lines in \css\ disfavor the possibility that the secondary star is a brown dwarf.

Indeed, at first blush, \css\ is almost identical to CSS~100603, the system
reported by \citet{breedt12}. In both of these sub-period-minimum systems, the
accretion disk shows high levels of both hydrogen and helium, and the late-type
donor is spectroscopically undetectable. In their theoretical examination of
AM~CVn progenitors, \citet{nelemans10} identify the presence of hydrogen as
conspicuous evidence favoring the evolved-CV channel over the double-white-dwarf
and helium-star channels of AM~CVn formation; if the secondary were either a
white dwarf or a helium star, hydrogen would likely be undetectable. Thus, both of these systems are excellent candidates for AM~CVn progenitors evolving pursuant to the evolved-main-sequence-donor model.

In neither system is the donor fully degenerate yet. To reach this conclusion about CSS~100603, \citet{breedt12} relied upon models of fully degenerate helium stars, and we took a similar approach with \css. Specifically, we used Equation 15 in \citet{vr88} (a mass-radius relation for a fully degenerate helium star) and set it equal to Equation 6 in \citet{knigge06} (a formula for the radius of a Roche-lobe-filling star given its mass and orbital period). Together, these formulae reveal that if the secondary were a degenerate helium star with an orbital period of 0.9227~hours, its mass would be 0.0083M$_{\odot}$---which, given the mass ratio of 0.06, implies an unreasonably low WD mass of 0.14M$_{\odot}$. Alternatively, if we adopt 0.83M$_{\odot}$ as the mass of the primary,\footnote{According to \citet{zsg11}, this is the average mass of primary stars in CVs.\label{mass}} the expected mass of the secondary would be 0.05M$_{\odot}$ according to the mass ratio. A helium star of this mass would have a radius of just 0.03R$_{\odot}$, making it considerably smaller than the radius of its Roche lobe (0.08R$_{\odot}$). Based on these considerations, we conclude that the donor in \css\ is semi-degenerate, and the system will continue to evolve toward the shorter orbital periods which characterize the majority of AM~CVn stars.

While it is likely that these two systems are the products of very similar
evolutionary processes, \css\ does have several features which differentiate it
from its cousin. The two most obvious dissimilarities are that \css\ is an
additional $\sim10$~minutes below the period minimum and has a mass ratio
$\sim$4~times greater than Breedt's CV. Moreover, \css\ shows more extensive
heavy-element enrichment, especially the Si~II 634.7/637.1-nm doublet and the
near-infrared Ca~II triplet, both of which are weak or non-existent in CSS~100603. As Section~\ref{disk_emission} explains, we also find that \css's disk is decidedly non-uniform---especially in the H$\alpha$ wavelength---and likely contains spiral structure. No comparable features have been reported in CSS~100603.

\subsection{Non-Uniform Disk Emission}\label{disk_emission}

\begin{figure}[hb!]
\epsscale{0.9}
\plotone{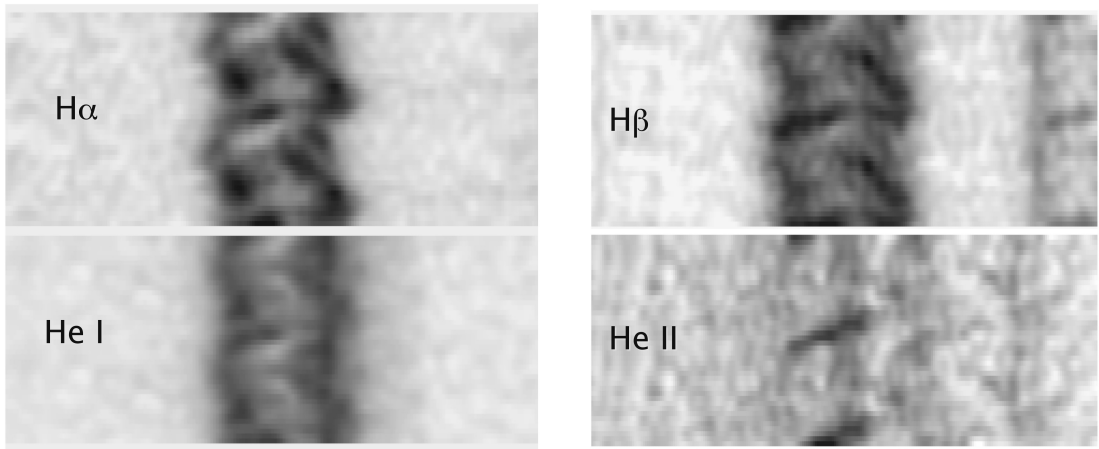}
\caption{\it{Trailed spectra of H$\alpha$, He~I 587.5~nm, H$\beta$, and He~II 468.6~nm. All trailed spectra show two orbits. To reduce contamination of the He~II line, the He~I 471.3~nm line has been partially subtracted using the profile of another, uncontaminated line.}
 \label{trailed_spectra}}
\end{figure}

The H$\alpha$ spectroscopy shows evidence of multiple emission regions on the disk. In the trailed H$\alpha$ spectra in Fig.~\ref{trailed_spectra}, the most prominent feature is the bright spot's classic S-wave, which oscillates between the H$\alpha$ line's two peaks over the course of the orbit. This S-wave vanishes as the bright spot moves from zero-velocity to maximum blueshift, only to reappear abruptly as the bright spot attains its maximum blueshift. During the bright spot's invisibility, a blueshifted absorption-like feature appears, but we suspect that it is the mere absence of emission rather than true absorption. In addition, near the maximum redshift of the bright spot, an even stronger emission feature appears in the blue wing of the H$\alpha$ line. As the bright spot transitions from maximum redshift to zero radial velocity, the blueshifted emission feature also moves toward zero velocity, but its intensity plummets dramatically.

\vspace{-0.2cm}

\begin{figure}[hb!]
\epsscale{.75}
\plotone{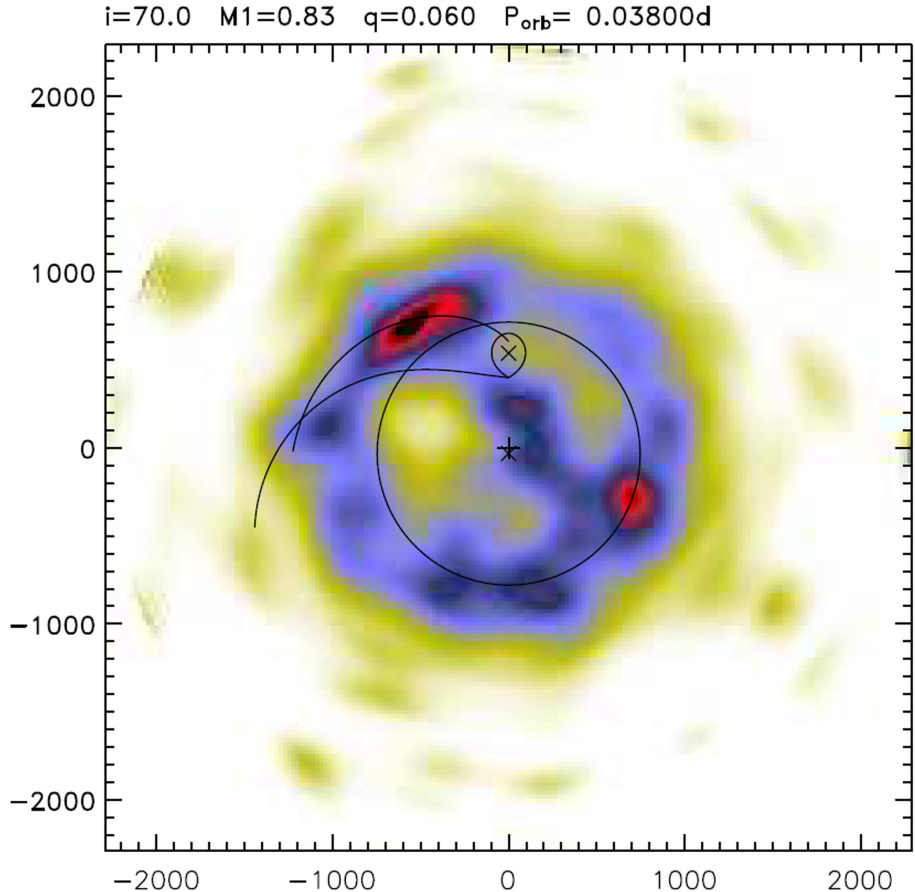}
\caption{\it{Doppler tomogram of the H$\alpha$ line. There are two distinct regions of emission, located on opposite sides of the disk. The tomogram plots the Roche lobe of the secondary assuming a mass ratio of $q = 0.06$, along with the accretion stream trajectory, the WD's position (denoted with an X), the center of mass (denoted with a + sign), and the orbital velocity of disk material at the 2:1 resonance as explained in Section \ref{spirals}.}
\label{tomogram}}
\end{figure} 

The appearance of the H$\beta$ and He~I lines (also in Fig.~\ref{trailed_spectra}) is somewhat less remarkable. The S-wave from the bright spot remains apparent, but the second emission feature is very subtle compared to the H$\alpha$ trailed spectrum. As with the H$\alpha$ line, the bright spot vanishes in these wavelengths as it moves from zero radial velocity to maximum blueshift.

The velocity information contained within a trailed spectrum can be used to reconstruct an indirect image of the disk in velocity coordinates (as opposed to spatial coordinates). Known as Doppler tomography, this technique essentially generates an inside-out image of the disk \citep{mh88}. Using the Doppler tomography algorithm of \citet{spruit}, we find that the competing S-waves in the H$\alpha$ trailed spectrum correspond with two distinct emission regions located on opposite sides of the disk (Fig.~\ref{tomogram}). The bright spot in the $-$V$_{x}$,+V$_{y}$ quadrant, which has a noticeably higher velocity than the other emission, is almost certainly attributable to the shock created by the stream-disk interaction, but the emission in the +V$_{x}$,$-$V$_{y}$ quadrant---which consists of both intermediate- and low-velocity components---presents a more of a mystery, one which we address in the following two subsections.

\subsubsection{Tidally-Induced Spiral Shocks}\label{spirals} In other CVs, there
have been at least two major proposed explanations for non-uniform emission of
this sort. One possibility is that the outer disk contains spiral arms produced
by the tidal torque of the secondary. Spiral shocks facilitate angular momentum
transfer in the disk, and in CVs with very low mass ratios, spirals can be permanent features in the disk because material in the outer disk attains a 2:1 orbital resonance with the donor \citep{lin79}. In CVs with relatively high mass ratios, tidal forces dissipate the outer disk before it can expand to the 2:1 resonance, but calculations by both \citet{lin79} and \citet{osaki02} found that when $q \lesssim 0.08$, the 2:1 resonance is inside the tidal truncation radius. Observationally, several CVs have shown strong evidence of spiral structure stemming from the 2:1 resonance \citep[e.g.][]{aviles10}.

We favor the presence of spiral shocks in as the source of the features observed in
our Doppler tomogram. Given that $q = 0.06$, the 2:1 resonance in \css\ rests within the tidal truncation radius, making it possible for permanent spiral structure to develop as described in \citet{lin79}. We further note that the overall appearance of our tomogram is very similar to a simulated tomogram of a disk with spiral arms caused by disk material at this very resonance \citep{aviles10}. To test more quantitatively for the presence of disk material at the 2:1 resonance, we again adopted a WD mass of 0.83M$_{\odot}$ (see our footnote~\ref{mass}) and computed the Keplerian velocity of disk material with an orbital period of exactly half of the system's orbital period. According to the formula $$v =\sqrt[3] {\frac{2\pi GM_{wd}}{P}},$$ the corresponding resonance velocity for these parameters is 750 km s$^{-1}$, which we plot as a circle around the origin in the tomogram. Although there are obvious pitfalls with simply assuming the WD's mass, the overlaid circle intersects much of the intense, non-uniform emission, furnishing circumstantial evidence that the disk in \css\ did extend to the 2:1 resonance when we observed the system.

Just as importantly, \citet{kunze05} calculated that spiral structure in such a system would produce a double-peaked light curve and presented a simulated phase plot which matches ours with uncanny accuracy. Kunze and Speith predict that both the minima and maxima will be unequal, with the deepest minimum coming at phase $\sim$0.25 and a weaker one at phase $\sim$0.75; the global maximum occurs at phase zero, and the weaker peak at phase $\sim$0.55.\footnote{Whereas we set phase zero equal to the peak of the light curve, Kunze and Speith selected the time of minimum flux as phase zero in their data. For the sake of internal consistency with the rest of this paper, we apply our own phasing scheme when describing their simulated phase plot.} Compared with the corresponding features in Kunze's and Speith's simulated light curve, the secondary maximum and minimum in our phase plot (Fig.~\ref{vatt}) are somewhat feeble in appearance. This relatively minor disparity might be because the simulations assume a higher inclination than the one actually observed in \css.

\subsubsection{Accretion-Stream Overflow}\label{stream_overflow} Alternatively,
non-uniform disk emission might be the result of an accretion stream which
overflows the disk after its initial collision, a mechanism which has received a
great deal of attention in theoretical studies \citep[e.g.,][]{al96, al98}. The
overflowing stream would cloak portions of the disk, producing absorption at
certain photometric phases and spectroscopic velocities, as it accelerated toward the inner disk. The reimpact of the stream with the disk, in turn, would produce an inner hotspot which, according to \citet{al98}, would be much more apparent in a system with a low-luminosity accretion disk. The emission in the +V$_{x}$,$-$V$_{y}$ quadrant of our tomogram, therefore, might be such a feature.

A shortcoming of the stream-overflow hypothesis is that the inner
bright spot should have a noticeably higher velocity than the outer bright spot
because the WD's strong gravity would significantly accelerate the overflowing
stream. We observe the reverse; the second bright spot in the +V$_{x}$,$-$V$_{y}$ quadrant has a {\it lower} velocity than the stream-disk interaction. Thus, the lower-than-expected velocity weighs against the possibility of an overflowing accretion stream. Furthermore, an overflowing stream would probably produce absorption at most phases, something which we do not observe. 

Though we cannot rule out the possibility of stream overflow, spiral shocks
elegantly weave the spectroscopy, photometry, and mass ratio of \css\ into a
reasonably coherent theory of the system, one which is less speculative than the
stream-overflow model.

\section{Conclusion}

We have reported photometry and spectroscopy of \cssfull, a CV with an orbital
period over 20~minutes below the period minimum. While the system is hydrogen-rich, its helium-to-hydrogen ratio is much higher than in typical SU~UMa-type
CVs, and the donor is completely invisible in our spectra. We identify
spectroscopic and photometric periods of 52.2~minutes and 55.36~minutes,
respectively. Using the 55-minute period and the previously reported superhump
period \citep{kato_astroph}, we estimate a mass ratio of $q =0.06$. Furthermore,
Doppler tomography reveals two distinct regions of intense H$\alpha$ emission on
the disk, consistent with spiral shocks produced when material in the outer disk
reaches a 2:1 resonance with the secondary. Drawing upon theoretical light
curves of low-mass-ratio CVs, we suspect that these spiral arms are responsible
for the intermittently double-peaked orbital modulation in the photometry, which
is reminiscent of the variation observed in WZ~Sge stars.

The best explanation for the short orbital period and the elevated helium
abundance is that \css\ is a progenitor of an AM~CVn system following the
evolved-CV track, similar to the system reported by \citet{breedt12}. The
discovery of two systems of this type in such rapid succession substantiates
theoretical predictions \citep[e.g.][]{nelemans10} that the evolved CV channel
of evolution can contribute significantly to the galactic AM CVn population.\footnote{\citet{breedt12} briefly mention two other sub-period-minimum CVs which could be AM~CVn progenitors with evolved donors, but at the time of writing, neither system has been studied in much detail.}


\acknowledgments

{\it Acknowledgments}

This paper has benefitted from the comments and suggestions of the two referees, to whom we are grateful.

 We thank Taichi Kato for sharing with us a power spectrum of his photometry of \css\ and helping us to rule out the 53-minute orbital period. 

R.K. received support from N.S.F. grant AST-1211196.

K.M. and A.A. received funding for this research through the Research Experience for Undergraduates program offered by the Department of Physics at the University of Notre Dame.

The results presented here are partially based on observations made with the VATT: the Alice P. Lennon Telescope and the Thomas J. Bannan Astrophysics Facility. We thank Richard Boyle and the Vatican Observatory Research Group for providing time on the VATT for this project.

The MODS spectrographs were built with funding from the NSF grant
AST-9987045 and the NSF Telescope System Instrumentation Program
(TSIP), with additional funds from the Ohio Board of Regents and the
Ohio State University Office of Research.

The LBT is an international collaboration among institutions in the United
States, Italy and Germany. LBT Corporation partners are: The University of Arizona on behalf of the Arizona university system; Istituto Nazionale di Astrofisica, Italy; LBT
Beteiligungsgesellschaft, Germany, representing the Max-Planck Society, the
Astrophysical Institute Potsdam, and Heidelberg University; The Ohio State
University, and The Research Corporation, on behalf of The University of Notre Dame,
University of Minnesota and University of Virginia.



{\it Facilities:} \facility{LBT, VATT, FLWO}

\clearpage

\clearpage

\begin{deluxetable}{lccc}
\tablecaption{Spectroscopy Log \label{speclog}}
\tablewidth{0pt}
\tablehead{
\colhead{Date} & \colhead{HJD} & \colhead{Exposure}  & \colhead{Instrument} \\
\colhead{(UT)} & \colhead{+2456000} & \colhead{(sec)} & \colhead{/Telescope} \\}
\startdata
2012 May 12 & $59.65\phantom{00}$ &  1800 & FAST/Tillinghast   \\
2012 May 13 & 60.6356 &  420 & FAST/Tillinghast   \\
2012 May 13 & 60.6412 &  420 & FAST/Tillinghast   \\
2012 May 13 & 60.6462 &  420 & FAST/Tillinghast   \\
2012 May 13 & 60.6512 &  420 & FAST/Tillinghast   \\
2012 May 13 & 60.6570 &  420 & FAST/Tillinghast   \\
2012 May 13 & 60.6620 &  420 & FAST/Tillinghast   \\
2012 May 13 & 60.6670 &  420 & FAST/Tillinghast   \\
2012 May 13 & 60.6727 &  420 & FAST/Tillinghast   \\
2012 May 13 & 60.6777 &  420 & FAST/Tillinghast   \\
2012 Jun 16 & 94.6645 &  400 & MODS/LBT   \\
2012 Jun 16 & 94.6680 &  400 & MODS/LBT   \\
2012 Jun 16 & 94.6716 &  400 & MODS/LBT   \\
2012 Jun 16 & 94.6751 &  400 & MODS/LBT   \\
2012 Jun 16 & 94.6786 &  400 & MODS/LBT   \\
2012 Jun 16 & 94.6821 &  400 & MODS/LBT   \\
2012 Jun 16 & 94.6856 &  400 & MODS/LBT   \\
2012 Jun 16 & 94.6891 &  400 & MODS/LBT   \\
2012 Jun 16 & 94.6926 &  400 & MODS/LBT   \\
2012 Jun 16 & 94.6962 &  400 & MODS/LBT   \\
2012 Jun 16 & 94.6997 &  400 & MODS/LBT   \\
2012 Jun 16 & 94.7032 &  400 & MODS/LBT   \\
2012 Jun 16 & 94.7068 &  400 & MODS/LBT   \\
2012 Jun 16 & 94.7103 &  400 & MODS/LBT   \\
2012 Jun 16 & 94.7138 &  400 & MODS/LBT   \\
2012 Jun 16 & 94.7173 &  400 & MODS/LBT   \\
\enddata
\end{deluxetable}

\end{document}